\documentclass{dsfe}

\usepackage{txfonts}
\usepackage{footmisc}

\def\typeofarticle{Research article}

\def\ppages{}
\def\DOI{}

\newtheorem{theorem}{Theorem}[section]
\newtheorem{lemma}[theorem]{Lemma}

\newtheorem{proposition}[theorem]{Proposition}

\newtheorem{corollary}[theorem]{Corollary}
\newtheorem{assumption}[theorem]{Assumption}

\numberwithin{equation}{section}

\newcommand{\e}{\mathrm{E}}
\newcommand{\p}{\mathrm{P}}
\newcommand{\var}{\mathrm{var}}

\begin{document}

\title{Proving prediction prudence}

\author{%
  Dirk Tasche\affil{1,}\corrauth
}

\shortauthors{Dirk Tasche}

\address{%
  \addr{\affilnum{1}}{Independent researcher}}

\corraddr{dirk.tasche@gmx.net.}

\begin{abstract}
We study how to perform tests on samples of pairs of observations and predictions
in order to assess whether or not the predictions are prudent. Prudence requires
that the mean of the difference of the observation-prediction pairs can be shown
to be significantly negative. For safe conclusions, we suggest testing 
both unweighted (or equally weighted) and weighted means and explicitly taking into account
the randomness of individual pairs. The test methods presented are mainly specified
as bootstrap and normal approximation algorithms. The tests are general but
can be applied in particular
in the area of credit risk, both for regulatory and accounting purposes.
\end{abstract}

\keywords{paired difference test; weighted mean; credit risk; probability of default (PD); 
loss given default (LGD); exposure at default (EAD); credit conversion factor (CCF)\\
\vspace{0.3cm}
\textbf{JEL Codes:} G21, C12}

\maketitle

\section{Introduction}
\label{se:intro}

Testing if the means of two samples significantly differ or the mean of one sample significantly 
exceeds the mean of the other sample is a problem that is widely covered in
the statistical literature \citep[see for instance][]{Casella&Berger, DavisonHinkley, venables2002modern}.
In this paper, we study how to perform such tests on samples of pairs of observations and predictions
in order to assess whether or not the predictions are prudent. Prudence is here understood 
as the requirement that
the mean of the differences of the observations and predictions can be shown
to be significantly negative.

At the latest by the validation requirements for credit risk parameter estimates 
in the regulatory Basel~II framework \citep[][paragraph~501]{BaselAccord},
such tests also became an important issue in the banking industry\footnote{%
PD means `probability of default', IRB means `internal ratings based', LGD means
`loss given default' and EAD is `exposure at default'.}: 
\begin{itemize}
 \item ``Banks must regularly compare realised default rates with estimated PDs for each
grade and be able to demonstrate that the realised default rates are within the expected
range for that grade'', and ``banks using the advanced IRB approach must complete such analysis
for their estimates of LGDs and EADs''.
\end{itemize}
More recently, as a consequence of the introduction of new rules for
loss provisioning in financial reporting standards, the validation of risk
parameter estimates also attracted interest in the accounting community 
\citep[see, e.g.,][]{bellini2019ifrs}.
Over the course of the past fifteen years or so, a variety of statistical 
tests for the comparison of realised and predicted values have been 
proposed for use in the banks' validation exercises. For overviews on estimation and
validation as well as 
references see \citet[][PD]{blumke2019out}, \citet[][LGD]{loterman2014proposed},
and \citet[][EAD]{gurtler2018exposure}. \citet{scandizzo2016validation} presents
validation methods for all these kinds of parameters in the general context of
model risk management.

In order to make 
validation results by different banks to some extent comparable, 
in February 2019, the European Central Bank \citep{ECBintructions}\footnote{%
In May 2020, this document could be downloaded at 
\url{https://www.bankingsupervision.europa.eu/banking/tasks/internal_models/shared/pdf/instructions_validation_reporting_credit_risk.en.pdf}.} asked the banks 
it supervises under the Single Supervisory Mechanism (SSM)
to deliver standardised annual reports on their internal model validation
exercises. In particular, the requested reports are assumed to include
data and tests regarding the ``predictive ability (or calibration)'' of PD, LGD and 
CCF (credit conversion factor)\footnote{%
EAD and CCF of a credit facility are linked by the relation $\text{EAD = DA + CCF*(limit-DA)}$
where DA is the already drawn amount.} parameters in the most recent observation period. 
Predictive ability for LGD estimation is explained through the statement 
``the analysis of predictive ability (or calibration) is aimed at ensuring 
that the LGD parameter adequately predicts the loss rate in the 
event of a default i.e.~that LGD estimates constitute reliable forecasts of realised loss rates''
\cite[][Section~2.6.2]{ECBintructions}. The meanings of predictive ability for PD and EAD / CCF 
respectively are illustrated in similar ways.

\citet{ECBintructions} proposed ``one-sample t-test[s] for paired observations'' 
to test the ``null hypothesis that estimated LGD [or CCF or EAD] is greater than true 
LGD'' (or CCF or EAD). \citet{ECBintructions} also suggested a Jeffreys binomial test
for the ``null hypothesis that the PD applied in the portfolio/rating grade at 
the beginning of the relevant observation period is greater than the true one 
(one sided hypothesis test)''. 

Recall that the possible outcomes of testing a null hypothesis against an alternative are
`the null hypothesis is not rejected' or `the null is rejected and the alternative is accepted'.
Not rejecting the null hypothesis does not mean accepting it because in hypothesis testing the
type II error (not rejecting the null hypothesis although the alternative is true) cannot be controlled
and, therefore, can be rather large. In contrast, the type I error (rejecting the null hypothesis although
it is true) can be controlled and usually is kept small by choosing a significance level like
5\% or 1\%. Hence, if the null hypothesis is rejected the alternative can be accepted at properly
controlled risk. In the following, we understand the acceptance of an alternative hypothesis by 
rejection of the null hypothesis as statistical `proof' with an error probability tag (i.e.\
the significance level or p-value).

In this paper,
\begin{itemize}
\item we make a case for also testing the null hypothesis that the estimated parameter is less than 
or equal to the true parameter in order to be able to `prove' that the estimate is
prudent (or conservative),
\item we suggest additionally using exposure- (or limit-)weighted\footnote{%
\citet{ECBintructions} presumably only looks at ``number-weighted''
(i.e.\ equally weighted) averages because the Basel framework 
\citep{BaselAccord} requires such averages for the risk parameter
estimates. In banking practice, however, also exposure-weighted 
averages are considered \citep[see, e.g.,][]{li2009validation}.} sample averages in order to 
better inform assessments of estimation (or prediction) prudence, and
\item we propose more elaborate statements of the
hypotheses for the tests (by including `variance expansion') 
in order to account for portfolio inhomogeneity in terms of
composition (exposure sizes) and riskiness.
\end{itemize}
The proposal to look for a `proof' of prediction prudence is inspired by
the regulatory requirement \citep[][paragraph~451]{BaselAccord}: 
``In order to avoid over-optimism, a bank must add to its estimates a margin of
conservatism that is related to the likely range of errors''.

As a matter of fact, the statistical tests discussed in this paper can be deployed 
both for proving prudence and for proving aggressiveness of estimates. However,
an unsymmetric approach is recommended for making use of the evidence from the tests:
\begin{itemize}
\item For proving prudence, request that both the equal-weights test and the exposure-weighted test reject
the null hypothesis of the parameter being aggressive.
\item For an alert of potential aggressiveness, request only that the equal-weights test or 
the exposure-weighted test reject the null hypothesis of the parameter being prudent.
\end{itemize}

The paper is organised as follows: 
\begin{itemize}
\item In Section~\ref{se:paired}, we introduce a general non-parametric
paired difference test approach to testing for the sign of a weighted mean value 
(Section~\ref{se:basic}). We compare this approach to the t-test for LGD, CCF and EAD proposed 
in \citet{ECBintructions} and note possible improvements of both approaches (Section~\ref{se:ECBttest}).
We then present in Section~\ref{se:LGD.CCF} a test approach to put into practice these 
improvements in the case of variables with values in the unit interval like LGD and CCF. 
Appendices~\ref{se:Cases} and \ref{se:EAD} supplement Section~\ref{se:LGD.CCF}
with regard to weight-adjustments as an alternative to sampling with inhomogeneous weights
and to testing non-negative but not necessarily bounded variables like EAD. 
\item In Section~\ref{se:PD}, we discuss paired difference tests in the special case
of differences between observed event indicators and the predicted probabilities of
the events. We start in Section~\ref{se:TestPD} with the presentation of a
test approach that takes account of potential weighting of the observation pairs 
and variance expansion to deal with the individual randomness of the observations.
In Section~\ref{se:ECBtest}, we compare this test approach to the Jeffreys test
proposed in \citet{ECBintructions} for assessing the `predictive ability' of
PD estimates.
\item In Section~\ref{se:example}, the test methods presented in the preceding sections are illustrated with
two examples of test results.
\item Section~\ref{se:concl} concludes the paper with summarising remarks.
\end{itemize}


\section{Paired difference tests}
\label{se:paired}

The statistical tests considered in this paper are `paired difference tests'.
This test design accounts for the strong dependence that is to be expected between
the observation and the prediction in the matched observation-prediction pairs 
which the analysed samples consist of.
See \citet[][Chapter~10]{mendenhall2008introduction} for a discussion of the
advantages of such test designs.

\subsection{Basic approach}
\label{se:basic}

\paragraph{Starting point.}\ 
\begin{itemize}
\item One sample of real-valued observations $\Delta_1, \ldots, \Delta_n$.
\item Weights $0 < w_i < 1$, $i = 1, \ldots, n$, with $\sum_{i=1}^n w_i = 1$.
\item Define the weighted-average observation $\Delta_w$ as
\begin{equation}\label{eq:Deltaw}
\Delta_w = \sum_{i=1}^n w_i\,\Delta_i.
\end{equation}
\end{itemize}

\begin{subequations}
\paragraph{Interpretation in the context of credit risk back-testing.} \ 
\begin{itemize}
\item $\Delta_1, \ldots, \Delta_n$ may be a sample of differences (residuals) 
between observed and predicted LGD (or CCF or EAD) for defaulted credit facilities (matched pairs
of observations and predictions).
\item The weight $w_i$ reflects the relative importance of observation $i$. For instance, in the case of 
CCF or EAD estimates of credit facilities, 
one might choose
\begin{equation}\label{eq:wCCF}
w_i \ = \ \frac{\text{limit}_i}{\sum_{j=1}^n \text{limit}_j},
\end{equation}
where $\text{limit}_j$ is the limit of credit facility $j$ at the time when the estimates were
made.
\item In case of LGD estimates, the weights $w_i$ could be chosen as \citep[][Section~5]{li2009validation}
\begin{equation}\label{eq:wLGD}
w_i \ = \ \frac{EAD_i}{\sum_{j=1}^n EAD_j},
\end{equation}
where $EAD_j$ is the exposure at default estimate for credit facility $j$ at the time when the estimates were
made.
\end{itemize}
\end{subequations}

\paragraph{Goal.} 
We consider $\Delta_w$ as defined by \eqref{eq:Deltaw} the
realisation of a test statistic to be defined below and want to answer the following two questions:
\begin{itemize}
\item If $\Delta_w <0$, how safe is the conclusion that the observed (realised) values are
on weighted average less than the predictions, i.e.\ the predictions are prudent / conservative?
\item If $\Delta_w > 0$, how safe is the conclusion that the observed (realised) values are
on weighted average greater than the predictions, i.e.\ the predictions are aggressive?
\end{itemize}
The safety of conclusions is measured by p-values which provide error probabilities for the
conclusions to be wrong. The lower the p-value, the more likely the conclusion is right.

In order to be able to examine the properties of 
the sample and $\Delta_w$ with statistical methods, 
we have to make the assumption that the sample was generated with
some random mechanism. 
The key idea for the mechanism is to interpret the weights $w_i$ as the probabilities 
of the corresponding observations $\Delta_i$. Consequently, we look at 
an inhomogeneous version of the empirical distribution of the sample $\Delta_1, \ldots, \Delta_n$, 
with the weight $w_i$ replacing $1/n$ as the probability of observation $\Delta_i$.
The details of the mechanism are described in the following assumption.

\begin{assumption}\label{as:X} 
The sample $\Delta_1, \ldots, \Delta_n$ consists of independent realisations of a random variable
$X_\vartheta$ with distribution given by
\begin{equation}\label{eq:DistrX}
\p\big[X_\vartheta = \Delta_i - \vartheta\big] \ =\ w_i, \qquad i = 1, \ldots, n,
\end{equation}
where the value of the parameter $\vartheta \in \mathbb{R}$ is unknown.
\end{assumption}
Note that \eqref{eq:DistrX} includes the case of equally weighted observations\footnote{%
See Appendix~\ref{se:Cases} for a more detailed discussion of special cases with
equal weights.}, by choosing $w_i = 1/n$ for all $i$.

\begin{proposition}\label{pr:moments}
For $X_\vartheta$ as described in Assumption~\ref{as:X}, the expected value and 
the variance are given by
\begin{subequations}
\begin{align}
\e[X_\vartheta] & = \Delta_w - \vartheta, \text{\ and}\label{eq:meanX}\\
\var[X_\vartheta]  & = \sum_{i=1}^n w_i\,\Delta_i^2 - \Delta_w^2.\label{eq:varX}
\end{align}
\end{subequations}
\end{proposition}

\textbf{Proof.} Obvious. \hfill\qed

By Assumption~\ref{as:X} and Proposition~\ref{pr:moments}, the questions on the
safety of conclusions from the sign of $\Delta_w$ can be translated into 
hypotheses on the value of the parameter $\vartheta$:
\begin{itemize}
\item If $\Delta_w < 0$, can we conclude that $H_0: \vartheta \le \Delta_w$ is false and
$H_1: \vartheta > \Delta_w \Leftrightarrow \e[X_\vartheta] <0$ is true?
\item If $\Delta_w > 0$, can we conclude that $H^\ast_0: \vartheta \ge \Delta_w$ is false and
$H^\ast_1: \vartheta < \Delta_w \Leftrightarrow \e[X_\vartheta] >0$ is true?
\end{itemize}

If we assume that the sample $\Delta_1, \ldots, \Delta_n$ was generated by independent
realisations of $X_\vartheta$ then the distribution of the sample mean is different from
the distribution of $X_\vartheta$, as shown in the following corollary to Proposition~\ref{pr:moments}.

\begin{corollary}\label{co:means}
Let $X_{1,\vartheta}, \ldots, X_{n,\vartheta}$ be independent and 
identically distributed copies of $X_\vartheta$ 
as in Assumption~\ref{as:X} and 
define $\bar{X}_\vartheta = \frac{1}{n} \sum_{i=1}^n X_{i,\vartheta}$. Then for 
the mean and the variance of $\bar{X}_\vartheta$, it holds that
\begin{subequations}
\begin{align}
\e[\bar{X}_\vartheta] & =  \Delta_w - \vartheta,\label{eq:meanXbar}\\
\var[\bar{X}_\vartheta]  & = \frac{1}{n} \Big(\sum_{i=1}^n w_i\,\Delta_i^2 - 
    \Delta_w^2\Big).\label{eq:var}
\end{align}
\end{subequations}
\end{corollary}
In the following, we use $\bar{X}_\vartheta$ as the test statistic and interpret $\Delta_w$ 
as its observed value\footnote{%
For arithmetic reasons, actually most of the time $\Delta_w$ cannot be a 
realisation of $\bar{X}_\vartheta$. As long as the sample size $n$ is not too small, however,
by \eqref{eq:meanXbar} and the law of large numbers considering $\Delta_w$ as
realisation of $\bar{X}_\vartheta$ is not unreasonable.}. 
Next we describe a bootstrap test to answer the above questions under Assumption~\ref{as:X}
and then provide the rationale behind its design.

\begin{subequations}
\paragraph{Bootstrap test.} Generate a Monte Carlo sample\footnote{%
According to \citet[][Section~5.2.3]{DavisonHinkley}, sample size $R=999$ should suffice
for the purposes of this paper.} 
$\bar{x}_1, \ldots, \bar{x}_R$ from $\Delta_1, \ldots, \Delta_n$ as follows:
\begin{itemize}
\item For $j=1, \ldots, R$: $\bar{x}_j$ is the equally weighted mean of $n$ 
independent draws from the distribution of $X_{\widehat{\vartheta}}$ as given by \eqref{eq:DistrX},
with $\widehat{\vartheta} = 0$. Equivalently,
$\bar{x}_j$ is the mean of $n$ draws with replacement from the sample $\Delta_1, 
\ldots, \Delta_n$, where $\Delta_i$ is drawn with probability $w_i$.
\item $\bar{x}_1, \ldots, \bar{x}_R$ are realisations of independent, 
identically distributed random variables.
\end{itemize}
Then a bootstrap p-value for the test of $H_0: \vartheta \le \Delta_w$ 
against $H_1: \vartheta > \Delta_w$
can be calculated as\footnote{%
$\#S$ denotes the number of elements of the set $S$.}
\begin{equation}\label{eq:bootp}
\text{p-value} \ = \ \frac{1 + \#\bigl\{i: i \in\{1, \ldots, n\}, 
    \bar{x}_i \le 2\,\Delta_w\bigr\}}{R+1}.
\end{equation}
A bootstrap p-value for the test of $H^\ast_0: \vartheta \ge \Delta_w$ 
against $H^\ast_1: \vartheta < \Delta_w$
is given by
\begin{equation}\label{eq:bootpStar}
\text{p-value}^\ast \ = \ \frac{1 + \#\bigl\{i: i \in\{1, \ldots, n\}, 
    \bar{x}_i \ge 2\,\Delta_w\bigr\}}{R+1}.
\end{equation}
\end{subequations}

\paragraph{Rationale.} By \eqref{eq:DistrX}, for each $\vartheta$ the
distributions of $X_0 - \vartheta$ and $X_\vartheta$ are identical.
As a consequence, if under $H_0$ the true parameter is $\vartheta \le \Delta_w$ and $(-\infty, \,x]$ is the 
critical (rejection) range for the test of $H_0$ against $H_1$ based 
on the test statistic $\bar{X}_\vartheta$, then it holds that
\begin{align}
\p\bigl[\bar{X}_\vartheta \in (-\infty, \,x]\bigr] &\ =\ \p[\bar{X}_0 \le x + \vartheta]\notag\\
 &\ \le \ \p[\bar{X}_0 \le x + \Delta_w].  \label{eq:delta}
\end{align}
Hence, by Theorem~8.3.27 of \citet{Casella&Berger}, 
in order to obtain a p-value for $H_0: \vartheta \le \Delta_w$ against $H_1: \vartheta > \Delta_w$, 
according to \eqref{eq:delta} it suffices to specify:
\begin{itemize}
 \item The upper limit $x$ of the critical range for rejection of $H_0: \vartheta \le \Delta_w$
 as `observed' value $\Delta_w$ of $\bar{X}_\vartheta$, and
 \item an approximation of the distribution of $\bar{X}_0$, as it is done by 
 generating the bootstrap sample $\bar{x}_1, \ldots, \bar{x}_R$.
\end{itemize} 
This implies Equation~\eqref{eq:bootp} for the bootstrap p-value\footnote{%
We adopt here the definition provided by \citet[][Eq.~(4.11)]{DavisonHinkley}.} of the test of $H_0$ 
against $H_1$. The rationale for \eqref{eq:bootpStar} is analogous.

\paragraph{Normal approximate test.} By Corollary~\ref{co:means} for $\vartheta=\Delta_w$, 
we find that the distribution of $\bar{X}_{\Delta_w}$ can be approximated by a normal distribution with 
mean 0 and variance as shown on the right-hand side of \eqref{eq:var}. With  $x= \Delta_w$,
therefore, we obtain the following expression 
for the normal approximate p-value of $H_0: \vartheta \le \Delta_w$ against $H_1: \vartheta > \Delta_w$:
\begin{subequations}
\begin{align}
\text{p-value} &\ = \ \p[\bar{X}_{\Delta_w} \le x] \notag\\
    & \ \approx\ \Phi\left(\frac{\sqrt{n}\,\Delta_w}
    {\sqrt{\sum_{i=1}^n w_i\,\Delta_i^2 - \Delta_w^2}}\right).\label{eq:normalp}
\end{align}
Here $\Phi$ denotes the standard normal distribution function.
The same reasoning gives for the normal approximate p-value of 
$H^\ast_0: \vartheta \ge \Delta_w$ against $H^\ast_1: \vartheta < \Delta_w$:
\begin{equation}\label{eq:normalpStar}
\text{p-value}^\ast  \ \approx\ 1 - \Phi\left(\frac{\sqrt{n}\,\Delta_w}
    {\sqrt{\sum_{i=1}^n w_i\,\Delta_i^2 - \Delta_w^2}}\right).
\end{equation}
\end{subequations}

\subsection{The t-test approach}
\label{se:ECBttest}

In Sections 2.6.2 (for LGD back-testing),
2.9.3.1 (for CCF back-testing) and
2.9.3.2 (for EAD back-testing) of \citet{ECBintructions}, the ECB  proposes 
a t-test for (in the terms of Section~\ref{se:basic} of this paper) $H^\ast_0: \vartheta \ge \Delta_w$ 
against $H^\ast_1: \vartheta < \Delta_w$.
Transcribed into the notation of Section~\ref{se:basic}, the test can
be described as follows:
\begin{itemize}
\item $n$ is the number of matched pairs of observations and predictions in the sample.
\item $\Delta_i$ is the difference of 
\begin{itemize}
\item the realised LGD for facility $i$ and the estimated LGD for facility $i$ in
    ECB Section~2.6.2,
\item the realised CCF for facility $i$ and the estimated CCF for facility $i$ in
   ECB  Section~2.9.3.1, and
\item the drawings (balance sheet exposure) at the time of default of facility $i$ and
    the estimated EAD of facility $i$ in ECB Section~2.9.3.2.
\end{itemize}
\item All $w_i$ equal $1/n$.
\item The right-hand side of \eqref{eq:var} is replaced by the sample variance
\begin{equation*}
s_n^2 \ = \ \frac{1}{n-1} \left(\frac{1}{n}\sum_{i=1}^n \Delta_i^2 - 
    \Delta_{1/n}^2\right).
\end{equation*}
\item The p-value is computed as
\begin{equation}\label{eq:pvaluet}
\text{p-value}^\ast \ = \ 1 - \Psi_{n-1}\left(\frac{\Delta_{1/n}}{s_n}\right),
\end{equation}
where $\Psi_{n-1}$ denotes the distribution function of Student's t-distribution
with $n-1$ degrees of freedom.
\end{itemize}
By the Central Limit Theorem, the p-values according to \eqref{eq:bootpStar}, \eqref{eq:normalpStar} and
\eqref{eq:pvaluet} will come out almost identical for large sample sizes $n$ and
equal weights $w_i = 1/n$ for all $i=1, \ldots, n$. For smaller $n$, 
the value of \eqref{eq:pvaluet} would be exact if the variables $X_{i, \vartheta}$ in 
Corollary~\ref{co:means} were normally distributed. 

\paragraph{Criticisms of the basic approach.} The basic approach as described in
Sections~\ref{se:basic} and \ref{se:ECBttest} fails to take account of the following issues:
\begin{itemize}
\item The random mechanism reflected by \eqref{eq:DistrX} can be interpreted as an expression of
uncertainty about the cohort / portfolio composition. The randomness of the loss rate / exposure
of the individual facilities -- the degree of which potentially can differ between facilities --
is not captured by \eqref{eq:DistrX}. 
\item The parametrisation of the distribution by a location parameter in \eqref{eq:DistrX} could
result in distributions with features that are not realistic, for instance negative exposures or loss rates
greater than one.
\end{itemize}
In the following section and in Appendix~\ref{se:EAD}, we are going to modify the basic approach for
LGD / CCF on the one hand and EAD on the other hand in such 
a way as to take into account these two issues. 

\subsection{Tests for variables with values in the unit interval}
\label{se:LGD.CCF}

By definition, both LGD and CCF take values only in the unit interval $[0,1]$. This fact allows
for more specific tests than the ones considered in the previous sections.
In this section, we talk only about 
LGD most of the time. But the concepts discussed also apply with little or no modification to
CCF or any other variables with values in the unit interval.

\paragraph{Starting point.}
\begin{itemize}
\item A sample of paired observations $(\lambda_1, \ell_1), \ldots, (\lambda_n, \ell_n)$, with 
predicted LGDs $0 < \lambda_i <1$ and 
realised loss rates $0 \le \ell_i \le 1$.
\item Weights $0 < w_i < 1$, $i = 1, \ldots, n$, with $\sum_{i=1}^n w_i = 1$,
\item Weighted average loss rate $\ell_w = \sum_{i=1}^n w_i\,\ell_i$ and
weighted average loss prediction $\lambda_w = \sum_{i=1}^n w_i\,\lambda_i$.
\end{itemize}

\paragraph{Interpretation in the context of LGD back-testing.} 
\begin{itemize}
\item A sample of $n$ defaulted credit facilities / loans is analysed.
\item The LGD $\lambda_i$ is an estimate of loan $i$'s loss rate as a consequence of the default, measured
as percentage of the exposure at the time of default (EAD).
\item The realized loss rate $\ell_i$ shows the percentage of loan $i$'s exposure at the time of default 
that cannot be recovered. 
\item The weight $w_i$ reflects the relative importance of observation $i$. In the case of LGD predictions, 
one might choose \eqref{eq:wLGD} for the definition of the weights, for CCF one might choose
\eqref{eq:wCCF} instead.
\item Define $\Delta_i = \ell_i - \lambda_i$, $i = 1, \ldots, n$. If $|\Delta_i| \approx 0$ then $\lambda_i$ 
is a good LGD prediction.
If $|\Delta_i| \approx 1$ then $\lambda_i$ is a poor LGD prediction.
\end{itemize}

\paragraph{Goal.} We want to use the observed weighted average difference / residual 
$\Delta_w = \sum_{i=1}^n w_i\,\Delta_i = \ell_w - \lambda_w$ to 
assess the quality of the calibration of the model / approach for the $\lambda_i$ to
predict the realised loss rates $\ell_i$. Again we want to answer the following two questions:
\begin{itemize}
\item If $\Delta_w <0$, how safe is the conclusion that the observed (realised) values are
on weighted average less than the predictions, i.e.\ the predictions are prudent / conservative?
\item If $\Delta_w > 0$, how safe is the conclusion that the observed (realised) values are
on weighted average greater than the predictions, i.e.\ the predictions are aggressive?
\end{itemize}
The safety of such conclusions is measured by p-values which provide error probabilities for the
conclusions to be wrong. The lower the p-value, the more likely the conclusion is right.

In order to be able to examine the specific properties of 
the sample and $\Delta_w$ with statistical methods, 
we have to make the assumption that the sample was generated with
some random mechanism. This mechanism is described in the following modification
of Assumption~\ref{as:X}.

\begin{subequations}
\begin{assumption}\label{as:LGD}
The sample $\Delta_1, \ldots, \Delta_n$ consists of independent realisations of a random variable
$X_\vartheta$ with distribution given by
\begin{equation}
X_\vartheta \ = \ \ell_I - Y_\vartheta,
\end{equation}
where $I$ is a random variable with values in $\{1, \ldots, n\}$ and $\p[I=i] = w_i$, $i=1, \ldots, n$.
$Y_\vartheta$ is a beta$(\alpha_i,\beta_i)$-distributed random variable\footnote{%
See \citet[][Section~3.3]{Casella&Berger} for a definition of the beta-distribution.} conditional
on $I = i$ for $i=1, \ldots, n$. 
The parameters $\alpha_i$ and $\beta_i$ of the beta-distribution depend on 
the unknown parameter $0 < \vartheta < 1$ by
\begin{equation}\label{eq:betapars}
\begin{split}
\alpha_i & \ = \ \vartheta_i\,\frac{1-v}{v}, \qquad\text{and}\\
\beta_i & \ = \ (1-\vartheta_i)\,\frac{1-v}{v}.
\end{split}
\end{equation}
In \eqref{eq:betapars}, the constant $0 < v < 1$ is the same for all $i$. The $\vartheta_i$ are
determined by
\begin{equation}\label{eq:theta.i}
\vartheta_i \ = \ (\lambda_i)^{h(\vartheta)},
\end{equation}
where $0 < h(\vartheta) < \infty$ is the unique solution $h$ of the equation
\begin{equation}
\vartheta \ = \ \sum_{i=1}^n w_i\,(\lambda_i)^h. 
\end{equation}
\end{assumption}
\end{subequations}

Assumption~\ref{as:LGD} introduces randomness of the difference between
loss rate and LGD prediction for individual facilities. Comparison 
between \eqref{eq:varLGD} below and \eqref{eq:varX} shows that 
this entails variance expansion of the sample $\Delta_1, \ldots, \Delta_n$.

Note that Assumption~\ref{as:LGD} also describes a method for
recalibration of the LGD estimates $\lambda_1, \ldots, \lambda_n$
to match targets $\vartheta$ with the weighted average of the $\vartheta_i$. In contrast to
\eqref{eq:DistrX}, the transformation \eqref{eq:theta.i} makes it sure that the transformed
LGD parameters still are values in the unit interval.
By definition of $Y_\vartheta$, it holds that $\e[Y_\vartheta\,|\,I=i] = \vartheta_i$. 

The constant $v$ specifies the variance of $Y_\vartheta$ conditional on $I=i$ as percentage of
the supremum $\vartheta_i\,(1-\vartheta_i)$ of its possible conditional variance, i.e.\ it holds that
\begin{equation}
\var[Y_\vartheta\,|\,I=i] \ = \ v\,\vartheta_i\,(1-\vartheta_i), \qquad i = 1, \ldots, n.
\end{equation}
The constant $v$ must be pre-defined or separately estimated. We suggest estimating it
from the sample $\ell_1, \ldots, \ell_n$ as 
\begin{equation}
\hat{v}\ =\ \frac{\sum_{i=1}^n w_i\,\ell_i^2 - \ell_w^2}{\ell_w\,(1-\ell_w)}.
\end{equation}
This approach yields $0 \le \hat{v} \le 1$ because the fact that $0 \le \ell_i \le 1$, $i = 1,
\ldots, n$, implies 
\begin{equation*}
\sum_{i=1}^n w_i\,\ell_i^2 - \ell_w^2\ \le\ \ell_w\,(1-\ell_w).
\end{equation*}

A simpler alternative to the definition \eqref{eq:theta.i} of $\vartheta_i$ would be linear scaling:
$\vartheta_i = \lambda_i\,\frac{\vartheta}{\lambda_w}$. However, with this definition
$\vartheta_i > 1$ may be incurred. This is not desirable because then the beta-distribution for
$Y_\vartheta\,|\,I=i$ would be ill-defined.

\begin{proposition}\label{pr:momentsLGD}
For $X_\vartheta$ as described in Assumption~\ref{as:LGD}, the expected value and 
the variance are given by
\begin{subequations}
\begin{align}
\e[X_\vartheta] & = \ell_w - \vartheta, \text{\ and}\label{eq:meanLGD}\\
\var[X_\vartheta]  & = \sum_{i=1}^n w_i\,(\ell_i-\vartheta_i)^2 - (\ell_w - \vartheta)^2 + 
    v \sum_{i=1}^n w_i\,\vartheta_i\,(1-\vartheta_i).\label{eq:varLGD}
\end{align}
\end{subequations}
\end{proposition}

\textbf{Proof.} For deriving the formula for $\var[X_\vartheta]$, make use of
the well-known variance decomposition \\[1ex]
\hspace*{3cm}$\displaystyle\var[X_\vartheta] = \e\bigl[\var[X_\vartheta\,|\,I]\bigr] +
\var\bigl[\e[X_\vartheta\,|\,I]\bigr]$.\hfill{\qed}

In contrast to \eqref{eq:varX}, the variance of $X_\vartheta$ as shown in \eqref{eq:varLGD}
depends on the parameter $\vartheta$ and has an additional 
component $v \sum_{i=1}^n w_i\,\vartheta_i\,(1-\vartheta_i)$
which reflects the potentially different variances of the loss rates in an inhomogeneous portfolio.

By Assumption~\ref{as:LGD} and Proposition~\ref{pr:momentsLGD}, the questions on the
safety of conclusions from the sign of $\Delta_w = \ell_w - \lambda_w$ again can be translated into 
hypotheses on the value of the parameter $\vartheta$:
\begin{itemize}
\item If $\Delta_w < 0$, can we conclude that $H_0: \vartheta \le \ell_w$ is false and
$H_1: \vartheta > \ell_w \Leftrightarrow \e[X_\vartheta] < 0$ is true?
\item If $\Delta_w > 0$, can we conclude that $H^\ast_0: \vartheta \ge \ell_w$ is false and
$H^\ast_1: \vartheta < \ell_w \Leftrightarrow \e[X_\vartheta]> 0$ is true?
\end{itemize}

If we assume that the sample $\Delta_1, \ldots, \Delta_n$ was generated by independent
realisations of $X_\vartheta$ then the distribution of the sample mean is different from
the distribution of $X_\vartheta$, as shown in the following corollary to Proposition~\ref{pr:momentsLGD}.

\begin{corollary}\label{co:meansLGD}
Let $X_{1, \vartheta}, \ldots, X_{n, \vartheta}$ be independent and 
identically distributed copies of $X_\vartheta$ 
as in Assumption~\ref{as:LGD} and 
define $\bar{X}_\vartheta = \frac{1}{n} \sum_{i=1}^n X_{i, \vartheta}$. Then for 
the mean and variance of $\bar{X}_\vartheta$, it holds that
\begin{subequations}
\begin{align}
\e[\bar{X}_\vartheta] & = \ell_w - \vartheta.\label{eq:meanLGDbar}\\
\var[\bar{X}_\vartheta]  & = \frac{1}{n} 
    \left(\sum_{i=1}^n w_i\,(\ell_i-\vartheta_i)^2 - (\ell_w - \vartheta)^2 + 
    v \sum_{i=1}^n w_i\,\vartheta_i\,(1-\vartheta_i)\right). \label{eq:varLGDbar}
\end{align}
\end{subequations}
\end{corollary}
In the following, we use $\bar{X}_\vartheta$ as the test statistic and interpret 
$\Delta_w = \ell_w - \lambda_w$ as its observed value.

\begin{proposition}\label{pr:LGD}
In the setting of Assumption~\ref{as:LGD} and Corollary~\ref{co:meansLGD}, 
$\vartheta \le \widehat{\vartheta}$ implies that
\begin{equation*}
\p[\bar{X}_\vartheta \le x]\ \le \ \p[\bar{X}_{\widehat{\vartheta}} \le x],\qquad \text{for all }x\in\mathbb{R}.
\end{equation*}
\end{proposition}
\textbf{Proof.} Observe that $\vartheta \le \widehat{\vartheta}$ implies 
$\vartheta_i \le \widehat{\vartheta}_i$ for all $i = 1, \ldots, n$.
For fixed $i$, the family of beta$(\alpha_i, \beta_i)$-distributions, parametrised by 
$\vartheta \in (0,1)$, has got a \emph{monotone likelihood ratio} in the sense
of Definition~8.3.16 of \citet{Casella&Berger}. This implies that for $\vartheta \le \widehat{\vartheta}$,
conditional on $I=i$, the distribution of $Y_{\widehat{\vartheta}}$ is stochastically 
not less than the distribution of $Y_{\vartheta}$, i.e.\ it holds that
\begin{equation*}
\p[Y_\vartheta \le x\,|\,I=i]\ \ge \ \p[Y_{\widehat{\vartheta}} \le x\,|\,I=i],\qquad \text{for all }x\in\mathbb{R}.
\end{equation*}
From this, it follows that for all $i= 1, \ldots,n$
\begin{equation*}
\p[X_\vartheta \le x\,|\,I=i]\ \le \ \p[X_{\widehat{\vartheta}} \le x\,|\,I=i],\qquad \text{for all }x\in\mathbb{R}.
\end{equation*}
But this inequality implies for all $x\in\mathbb{R}$ that 
\begin{equation}\label{eq:smaller}
\p[X_\vartheta \le x]\ = \ \sum_{i=1}^n w_i\, \p[X_\vartheta \le x\,|\,I=i] \ \le \ \p[X_{\widehat{\vartheta}} \le x].
\end{equation}
Property \eqref{eq:smaller} is passed on to convolutions of independent copies of $X_\vartheta$ and 
$X_{\widehat{\vartheta}}$. This proves the assertion. 
\hfill \qed

\begin{subequations}
\paragraph{Bootstrap test.} Generate a Monte Carlo sample
$\bar{x}_1, \ldots, \bar{x}_R$ from $X_\vartheta$ with $\vartheta = \ell_w$ as follows:
\begin{itemize}
\item For $j=1, \ldots, R$: $\bar{x}_j$ is the equally weighted mean of $n$ 
independent draws from the distribution of $X_{\vartheta}$ as given by Assumption~\ref{as:LGD},
with $\vartheta = \ell_w$. 
\item $\bar{x}_1, \ldots, \bar{x}_R$ are realisations of independent, 
identically distributed random variables.
\end{itemize}
Then a bootstrap p-value for the test of $H_0: \vartheta \le \ell_w$ 
against $H_1: \vartheta > \ell_w$ can be calculated as
\begin{equation}\label{eq:bootpLGD}
\text{p-value} \ = \ \frac{1 + \#\bigl\{i: i \in\{1, \ldots, n\}, 
    \bar{x}_i \le \ell_w- \lambda_w\bigr\}}{R+1}.
\end{equation}
A bootstrap p-value for the test of $H^\ast_0: \vartheta \ge \ell_w$ 
against $H^\ast_1: \vartheta < \ell_w$ is given by
\begin{equation}\label{eq:bootpStarLGD}
\text{p-value}^\ast \ = \ \frac{1 + \#\bigl\{i: i \in\{1, \ldots, n\}, 
    \bar{x}_i \ge \ell_w- \lambda_w\bigr\}}{R+1}.
\end{equation}
\end{subequations}

\paragraph{Rationale.} By Proposition~\ref{pr:LGD}, 
if under $H_0$ the true parameter is $\vartheta \le \ell_w$ and $(-\infty, \,x]$ is the 
critical (rejection) range for the test of $H_0: \vartheta \le \ell_w$ against $H_1: \vartheta > \ell_w$ based 
on the test statistic $\bar{X}_\vartheta$, then it holds that
\begin{equation}
\p\bigl[\bar{X}_\vartheta \in (-\infty, \,x]\bigr] \ \le\ 
    \p[\bar{X}_{\ell_w} \le x]. \label{eq:xLGD}
\end{equation}
Hence, by Theorem~8.3.27 of \citet{Casella&Berger}, 
in order to obtain a p-value for $H_0: \vartheta \le \ell_w$ against $H_1: \vartheta >\ell_w$, 
according to \eqref{eq:xLGD} it suffices to specify:
\begin{itemize}
 \item The upper limit $x$ of the critical range for rejection of $H_0: \vartheta \le \ell_w$
 as our realisation $\Delta_w =\ell_w- \lambda_w$ of $\bar{X}_\vartheta$, and
 \item an approximation of the distribution of $\bar{X}_{\ell_w}$, as it has been done by 
 generating the bootstrap sample $\bar{x}_1, \ldots, \bar{x}_R$.
\end{itemize} 
This implies Equation~\eqref{eq:bootpLGD} for the bootstrap p-value. 
The rationale for \eqref{eq:bootpStarLGD} is analogous.

\begin{subequations}
\paragraph{Normal approximate test.} By Corollary~\ref{co:meansLGD}, 
we find that the distribution of $\bar{X}_{\ell_w}$ can be approximated by a normal distribution with 
mean 0 and variance as shown  on the right-hand side of \eqref{eq:varLGD} with $\vartheta=\ell_w$. 
With  $x=\ell_w-\lambda_w$,
one obtains for the approximate p-value of $H_0: \vartheta \le \ell_w$ against $H_1: \vartheta > \ell_w$:
\begin{align}
\text{p-value} &\ = \ \p[\bar{X}_{\ell_w} \le x] \notag\\
    & \ \approx\ \Phi\left(\frac{\sqrt{n}\,(\ell_w - \lambda_w)}
    {\sqrt{\sum_{i=1}^n w_i\,(\ell_i-\widehat{\vartheta}_i)^2  + 
    v\,\sum_{i=1}^n w_i\,\widehat{\vartheta}_i\,(1-\widehat{\vartheta}_i)}}\right),\label{eq:normalpLGD}
\end{align}
with $\widehat{\vartheta}_i = (\lambda_i)^{h(\ell_w)}$ as in
Assumption~\ref{as:LGD}.
The same reasoning gives for the normal approximate p-value of 
$H^\ast_0: \vartheta \ge \ell_w$ against $H^\ast_1: \vartheta < \ell_w$:
\begin{equation}\label{eq:normalpStarLGD}
\text{p-value}^\ast  \ \approx\ 1 - \Phi\left(\frac{\sqrt{n}\,(\ell_w - \lambda_w)}
    {\sqrt{\sum_{i=1}^n w_i\,(\ell_i-\widehat{\vartheta}_i)^2  + 
    v\,\sum_{i=1}^n w_i\,\widehat{\vartheta}_i\,(1-\widehat{\vartheta}_i)}}\right).
\end{equation}
\end{subequations}


\section{Tests of probabilities}
\label{se:PD}

\paragraph{Starting point.}\ 
\begin{itemize}
\item A sample of paired observations $(p_1, b_1), \ldots, (p_n, b_n)$, 
with probabilities $0 < p_i <1$ and status indicators $b_i \in \{0, 1\}$ (1 for defaulted, 0 for
performing).
\item Weights $0 < w_i < 1$, $i = 1, \ldots, n$, with $\sum_{i=1}^n w_i = 1$,
\item Weighted default rate $b_w = \sum_{i=1}^n w_i\,b_i$ and weighted 
average PD $p_w = \sum_{i=1}^n w_i\,p_i$.
\end{itemize}

\textbf{Interpretation in the context of PD back-testing.}\  
\begin{itemize}
\item A sample of $n$ borrowers is observed for a certain period of time, most commonly one year.
\item The PD $p_i$ is an estimate of borrower $i$'s probability to default during the observation period, estimated
before the beginning of the period.
\item The status indicator $b_i$ shows borrower $i$'s performance status 
at the end of the observation period. $b_i =1$ means
``borrower has defaulted'', $b_i = 0$ means ``borrower is performing''.
\item $w_i$ could be the relative importance of observation $i$. In the case of default predictions, 
one might choose weights as in \eqref{eq:wLGD}.
\item Define $\Delta_i = b_i - p_i$, $i = 1, \ldots, n$. If $|\Delta_i| \approx 0$ 
then $p_i$ is a good default prediction.
If $|\Delta_i| \approx 1$ then $p_i$ is a poor default prediction.
\end{itemize}

\paragraph{Goal.} We want to use the observed weighted average difference / residual 
$\Delta_w = \sum_{i=1}^n w_i\,\Delta_i = b_w - p_w$ to 
assess the quality of the calibration of the model / approach for the $p_i$ to
predict the realised status indicators $b_i$. Again we want to answer the following two questions:
\begin{itemize}
\item If $\Delta_w <0$, how safe is the conclusion that the observed (realised) values are
on weighted average less than the predictions, i.e.\ the predictions are prudent / conservative?
\item If $\Delta_w > 0$, how safe is the conclusion that the observed (realised) values are
on weighted average greater than the predictions, i.e.\ the predictions are aggressive?
\end{itemize}
The safety of such conclusions is measured by p-values which provide error probabilities for the
conclusions to be wrong. The lower the p-value, the more likely the conclusion is right.
In determining the p-values, we take into account the criticisms of the basic approach
as mentioned at the end of Section~\ref{se:ECBttest}.

\subsection{Testing probabilities on inhomogeneous samples}
\label{se:TestPD}

In order to be able to examine the PD-specific properties of 
the sample and $\Delta_w =b_w -p_w$ with statistical methods, 
we have to make the assumption that the sample was generated with
some random mechanism. This mechanism is described in the following modification
of Assumptions~\ref{as:X} and \ref{as:LGD}.

\begin{subequations}
\begin{assumption}\label{as:PD}
The sample $\Delta_1, \ldots, \Delta_n$ consists of independent realisations of a random variable
$X_\vartheta$ with distribution given by
\begin{equation}
X_\vartheta \ = \ b_I - Y_\vartheta,
\end{equation}
where $I$ is a random variable with values in $\{1, \ldots, n\}$ and $\p[I=i] = w_i$, $i=1, \ldots, n$.
$Y_\vartheta$ is a Bernoulli variable with 
\begin{equation}
\p[Y_\vartheta=1\,|\,I=i] \ = \ \vartheta_i, \qquad i=1, \ldots, n.
\end{equation}
Define $\varrho_i = \frac{1-p_i}{p_i}\,\frac{p_w}{1-p_w}$. Then the $\vartheta_i$ 
depend on the unknown parameter $0 < \vartheta < 1$ by
\begin{equation}\label{eq:thetaPD}
\vartheta_i = \frac{\vartheta}{\vartheta +(1-\vartheta)\,\varrho_i\,h(\vartheta)},
\end{equation}
where $0< h(\vartheta)<\infty$ is the unique\footnote{%
See \citet[][Section~4.2.4]{tasche2013art}.} solution of the equation 
\begin{equation}\label{eq:Defh}
1\ = \ \sum_{i=1}^n \frac{w_i}{\vartheta +(1-\vartheta)\,\varrho_i\,h},
\end{equation}
when solved for $h$.
\end{assumption}
\end{subequations}

Assumption~\ref{as:PD} introduces randomness of the difference between
status indicator and PD prediction for individual facilities. Comparison 
between \eqref{eq:varPD} below and \eqref{eq:varX} shows that 
this entails variance expansion of the sample $\Delta_1, \ldots, \Delta_n$.

Note that Assumption~\ref{as:PD} also describes a method for
recalibration of the PD estimates $p_1, \ldots, p_n$
to match targets $\vartheta$ with the weighted average of the $\vartheta_i$. In contrast to
\eqref{eq:DistrX}, the transformation \eqref{eq:thetaPD} makes it sure that the transformed
PD parameters still are values in the unit interval. In principle, instead 
of \eqref{eq:thetaPD} also the transformation \eqref{eq:theta.i} could have been used.
\eqref{eq:thetaPD} was preferred because it has a probabilistic foundation through Bayes' theorem.
By definition of $Y_\vartheta$, it holds that $\e[Y_\vartheta\,|\,I=i] = \vartheta_i$. 

Another simple alternative to the definition \eqref{eq:thetaPD} of $\vartheta_i$ would be linear scaling:
$\vartheta_i = p_i\,\frac{\vartheta}{p_w}$. However, with this definition
$\vartheta_i > 1$ may be incurred. This is not desirable because then the Bernoulli distribution for
$Y_\vartheta\,|\,I=i$ would be ill-defined.

\begin{proposition}\label{pr:momentsPD}
For $X_\vartheta$ as described in Assumption~\ref{as:PD}, 
the expected value and the variance are given by
\begin{subequations}
\begin{align}
\e[X_\vartheta] & = b_w - \vartheta, \text{\ and}\label{eq:meanPD}\\
\var[X_\vartheta]  & = \sum_{i=1}^n w_i\,(b_i-\vartheta_i)^2 - (b_w - \vartheta)^2 + 
    \sum_{i=1}^n w_i\,\vartheta_i\,(1-\vartheta_i).\label{eq:varPD}
\end{align}
\end{subequations}
\end{proposition}

\textbf{Proof.} Similar to the proof of Proposition~\ref{pr:momentsLGD}. \hfill\qed

Note that $\sum_{i=1}^n w_i\,(b_i-\vartheta_i)^2$ is a weighted version of the 
Brier Score \citep[see, e.g.,][]{Hand97} for the
observation-prediction sample $(b_1, \vartheta_i), \ldots, (b_n, \vartheta_n)$. 
This observation suggests that the power of the calibration tests
considered in this section will be the greater, the better the discriminatory power
of the PD predictions is (reflected by lower Brier scores).

By Assumption~\ref{as:PD} and Proposition~\ref{pr:momentsPD}, the questions on the
safety of conclusions from the sign of $\Delta_w = b_w - p_w$ again can be translated into 
hypotheses on the value of the parameter $\vartheta$:
\begin{itemize}
\item If $\Delta_w < 0$, can we conclude that $H_0: \vartheta \le b_w$ is false and
$H_1: \vartheta > b_w \Leftrightarrow \e[X_\vartheta] < 0$ is true?
\item If $\Delta_w > 0$, can we conclude that $H^\ast_0: \vartheta \ge b_w$ is false and
$H^\ast_1: \vartheta < b_w \Leftrightarrow \e[X_\vartheta]> 0$ is true?
\end{itemize}

If we assume as before in Section~\ref{se:paired} that the sample $\Delta_1, \ldots, \Delta_n$ 
was generated by independent
realisations of $X_\vartheta$ then the distribution of the sample mean is different from
the distribution of $X_\vartheta$, as shown in the following corollary to Proposition~\ref{pr:momentsPD}.

\begin{corollary}\label{co:meansPD}
Let $X_{1, \vartheta}, \ldots, X_{n, \vartheta}$ be independent and 
identically distributed copies of $X_\vartheta$ 
as in Assumption~\ref{as:PD} and 
define $\bar{X}_\vartheta = \frac{1}{n} \sum_{i=1}^n X_{i, \vartheta}$. Then for 
the mean and variance of $\bar{X}_\vartheta$, it holds that
\begin{subequations}
\begin{align}
\e[\bar{X}_\vartheta] & = b_w - \vartheta.\label{eq:meanPDbar}\\
\var[\bar{X}_\vartheta]  & = \frac{1}{n} \left(\sum_{i=1}^n w_i\,(b_i-\vartheta_i)^2 - (b_w - \vartheta)^2 + 
    \sum_{i=1}^n w_i\,\vartheta_i\,(1-\vartheta_i)\right). \label{eq:varPDbar}
\end{align}
\end{subequations}
\end{corollary}

In the following, we use $\bar{X}_\vartheta$ as the test statistic and interpret 
$\Delta_w = b_w - p_w$ as its observed value.

\begin{lemma}\label{le:PD}
In the setting of Assumption~\ref{as:PD}, $\vartheta < \widehat{\vartheta}$ implies that
$\vartheta_i < \widehat{\vartheta}_i$ for all $i = 1, \ldots, n$.
\end{lemma}
\textbf{Proof.} Assume $\vartheta < \widehat{\vartheta}$ and let $h = h(\vartheta)$ and
$\widehat{h} = h\bigl(\widehat{\vartheta}\bigr)$. Along the same lines of algebra as in Section~3
of \citet{tasche2013law}, it can be shown that (with $w_i$ and $\varrho_i$ as in Assumption~\ref{as:PD}) for
$0 < t < 1$ and $\eta >0$ the following two equations are equivalent:
\begin{equation}\label{eq:equiv}
\begin{split}
1\ = \ \sum_{i=1}^n \frac{w_i}{t +(1-t)\,\varrho_i\,\eta} \\
\iff \quad 0 \ = \ \sum_{i=1}^n \frac{w_i\,(1-\varrho_i\,\eta)}{t +(1-t)\,\varrho_i\,\eta}.
\end{split}
\end{equation}
Define $f(t, \eta) = \sum_{i=1}^n \frac{w_i\,(1-\varrho_i\,\eta)}{t +(1-t)\,\varrho_i\,\eta}$. Then
we obtain
\begin{subequations}
\begin{align}
\frac{\partial f}{\partial t}(t, \eta) &\ =\ - \sum_{i=1}^n \frac{w_i\,(1-\varrho_i\,\eta)^2}
    {(t +(1-t)\,\varrho_i\,\eta)^2}\ < \ 0, \label{eq:tpartial}\\
\frac{\partial f}{\partial \eta}(t, \eta) &\ =\ - \sum_{i=1}^n \frac{w_i\,\varrho_i}
    {(t +(1-t)\,\varrho_i\,\eta)^2}\ < \ 0.\label{eq:etapartial}
\end{align}
\end{subequations}
By definition, \eqref{eq:Defh} holds for $\vartheta$ and $h$. From \eqref{eq:equiv} and \eqref{eq:tpartial}
then it follows that
\begin{equation*}
0\ > \ \sum_{i=1}^n \frac{w_i\,(1-\varrho_i\,h)}{\widehat{\vartheta} +(1-\widehat{\vartheta})\,\varrho_i\,h}.
\end{equation*}
However, by \eqref{eq:equiv} we also have
\begin{equation*}
0\ = \ \sum_{i=1}^n \frac{w_i\,(1-\varrho_i\,\widehat{h})}{\widehat{\vartheta} +
    (1-\widehat{\vartheta})\,\varrho_i\,\widehat{h}}.
\end{equation*}
By \eqref{eq:etapartial}, this only is possible if it holds that $h > \widehat{h}$. Hence it follows that
\begin{equation*}
\frac{(1-\vartheta)\,h}{\vartheta} \ > \ \frac{(1-\widehat{\vartheta})\,\widehat{h}}{\widehat{\vartheta}}.
\end{equation*}
By \eqref{eq:thetaPD} (i.e.~the definition of $\vartheta_i$ and $\widehat{\vartheta}_i$), this inequality implies
$\vartheta_i < \widehat{\vartheta}_i$. \hfill \qed

\begin{theorem}\label{th:PD}
In the setting of Assumption~\ref{as:PD} and Corollary~\ref{co:meansPD}, 
$\vartheta \le \widehat{\vartheta}$ implies that
\begin{equation*}
\p[\bar{X}_\vartheta \le x]\ \le \ \p[\bar{X}_{\widehat{\vartheta}} \le x],\qquad \text{for all }x\in\mathbb{R}.
\end{equation*}
\end{theorem}
\textbf{Proof.} By Lemma~\ref{le:PD}, $\vartheta \le \widehat{\vartheta}$ implies for all $i= 1, \ldots,n$ that
$\vartheta_i \le \widehat{\vartheta}_i$ and therefore also 
\begin{equation*}
\p[Y_\vartheta \le x\,|\,I=i]\ \ge \ \p[Y_{\widehat{\vartheta}} \le x\,|\,I=i],\qquad \text{for all }x\in\mathbb{R}.
\end{equation*}
The remainder of the proof is identical to the last part of the proof of 
Proposition~\ref{pr:LGD}.
\hfill{\qed}

\paragraph{Exact p-values.} Since by definition up to the constant $1/n$ 
the test statistic $\bar{X}_\vartheta$ as defined in Assumption~\ref{as:PD} 
and Corollary~\ref{co:meansPD} takes only integer values in the 
range $\{-n, \ldots, -1, 0, 1, \ldots, n\}$, its distribution 
can readily be exactly determined by means of an inverse Fourier transform
\citep[][Section~4.7]{Rolski&al}. 
By Theorem~\ref{th:PD} and Theorem~8.3.27 of 
\citet{Casella&Berger}, then a p-value for the test of $H_0: \vartheta \le b_w$ 
against $H_1: \vartheta > b_w$ can exactly be computed as
\begin{subequations}
\begin{equation}\label{eq:p}
\text{p-value}\ = \ \p[\bar{X}_{b_w} \le b_w - p_w].
\end{equation}
A p-value for the test of $H^\ast_0: \vartheta \ge b_w$ 
against $H^\ast_1: \vartheta < b_w$ is given by
\begin{equation}\label{eq:pStar}
\text{p-value}^\ast \ = \ \p[\bar{X}_{b_w} \ge b_w - p_w].
\end{equation}
\end{subequations}

\begin{subequations}
\paragraph{Normal approximate test.} By Corollary~\ref{co:meansPD}, 
we find that the distribution of $\bar{X}_{b_w}$ can be approximated by a normal distribution with 
mean 0 and variance as shown  on the right-hand side of \eqref{eq:varPDbar}. With  $x=b_w-p_w$,
one obtains for the approximate p-value of $H_0: \vartheta \le b_w$ against $H_1: \vartheta > b_w$:
\begin{align}
\text{p-value} &\ = \ \p[\bar{X}_{b_w} \le x] \notag\\
    & \ \approx\ \Phi\left(\frac{\sqrt{n}\,(b_w - p_w)}
    {\sqrt{\sum_{i=1}^n w_i\,(b_i-\widehat{\vartheta}_i)^2  + 
    \sum_{i=1}^n w_i\,\widehat{\vartheta}_i\,(1-\widehat{\vartheta}_i)}}\right),\label{eq:normalpProb}
\end{align}
with $\widehat{\vartheta}_i = \frac{b_w}{b_w +(1-b_w)\,\varrho_i\,h(b_w)}$ as in
Assumption~\ref{as:PD}.
The same reasoning gives for the normal approximate p-value of 
$H^\ast_0: \vartheta \ge \ell_w$ against $H^\ast_1: \vartheta < \ell_w$:
\begin{equation}\label{eq:normalpStarProb}
\text{p-value}^\ast  \ \approx\ 1 - \Phi\left(\frac{\sqrt{n}\,(b_w - p_w)}
    {\sqrt{\sum_{i=1}^n w_i\,(b_i-\widehat{\vartheta}_i)^2  + 
    \sum_{i=1}^n w_i\,\widehat{\vartheta}_i\,(1-\widehat{\vartheta}_i)}}\right).
\end{equation}
\end{subequations}

\subsection{The Jeffreys test approach}
\label{se:ECBtest}

In Section~2.5.3.1 of \citet{ECBintructions}, the ECB  proposes 
``PD back testing using a Jeffreys test''. 
Transcribed into the notation of Section~\ref{se:TestPD} of this paper, the starting point
for the test can be described as follows:
\begin{itemize}
\item $n=N$, where ``N is the number of customers in the portfolio/rating grade''.
\item $\sum_{i=1}^n b_i = D$, where ``D is the number of those customers that 
    have defaulted within that observation period''.
\item $\frac{1}{n}\sum_{i=1}^n p_i =PD$, where PD means the ``PD [probability of default] 
    of the portfolio/rating grade''.
\item All $w_i$ equal $1/n$.
\end{itemize}

\paragraph{The Jeffreys test for the success parameter of a binomial distribution.}\ 
\begin{itemize}
 \item In a Bayesian setting, an ``objective Bayesian'' prior distribution beta$(1/2, 1/2)$
for the PD 
is chosen such that -- assuming a binomial distribution for the number of defaults -- the
posterior distribution (i.e.\ conditional on the observed number of defaults) of the PD is
beta$(D+1/2, \,N-D+1/2)$. See \citet{Kazianka2016} for the rationale for choosing this 
method of test. If estimated as the mean of the posterior distribution, the Bayesian PD
estimate is $\frac{D+1/2}{N+1}$.
\item The Null hypothesis is ``the PD applied in the portfolio/rating grade $\ldots$ is greater than the true one 
(one sided hypothesis test)'', i.e.\ $H_0: \theta \le \widehat{\theta}$ with $\widehat{\theta} =$ 
``applied PD'' and $\theta =$ ``true PD''. In the notation of Section~\ref{se:TestPD}, this can
be phrased as testing $H^\ast_0: \vartheta \ge b_{1/n}$ against $H^\ast_1: \vartheta < b_{1/n}$.
\item \citet{ECBintructions}: ``The test statistic is the PD of the portfolio/rating grade.''
The construction principle for the Jeffreys test is to determine a credibility interval for
the PD and then to check if the applied PD is inside or outside of the interval.
\item The p-value for this kind of Jeffreys test is 
\begin{equation}\label{eq:Jeffreys}
\text{p-value}_{\text{Jeffreys}} \ =\ F_{D+1/2,\,N-D+1/2}(PD),
\end{equation}
where $F_{\alpha,\,\beta}$  denotes the distribution function of the beta$(\alpha, \beta)$-distribution.
\end{itemize}
\paragraph{Comments.}\ 
\begin{itemize}
\item The standard (frequentist) one-sided binomial test would be: `Reject $H_0$ if $D \ge c$' where $c$ is 
a `critical' value such the probability under $H_0$ to observe $c$ or more defaults is small. 
For this test, the p-value is
\begin{equation}\label{eq:freq}
\text{p-value}_{\text{freq}}\ = \ \sum_{i=D}^N \bigl(\begin{smallmatrix} N \\ i\end{smallmatrix}\bigr)\,
    PD^i\,(1-PD)^{N-i} \ = \ F_{D,\,N-D+1}(PD).
\end{equation}
Hence, unless the observed number of default $D$ is very small or even zero, from \eqref{eq:Jeffreys} it
follows that in practice most of the time the Jeffreys test and the standard binomial test give
similar results.
\item For a `fair' comparison of the Jeffreys test and the test proposed in Section~\ref{se:TestPD},
we have to modify Assumption~\ref{as:PD} such that there is no variance expansion and all
weights are equal, i.e.\ the random
variable $X_\vartheta$ is simply defined by
\begin{equation}\label{eq:Xsimple}
\p[X_\vartheta = b_i - \vartheta_i] \ = \ \frac{1}{n},\qquad i = 1, \ldots, n,
\end{equation}
where the $\vartheta_i$ depend on the unknown parameter $0 < \vartheta < 1$ in the way 
described by \eqref{eq:thetaPD} and \eqref{eq:Defh}. The normal approximate p-value
of $H_0$ against $H_1$ is then (using the ECB notation) 
\begin{equation}\label{eq:simple}
\text{p-value} \ \approx\ 1 - \Phi\left(\frac{\sqrt{N}\,(D/N - PD)}
    {\sqrt{D/N\,(1-D/N)}}\right).
\end{equation}
\item The normal approximation of the frequentist (and by \eqref{eq:Jeffreys} and 
    \eqref{eq:freq} also Jeffreys) binomial test p-value is
\begin{equation}\label{eq:freqNorm}
\text{p-value}_{\text{freq}}\ \approx \ 
    1 - \Phi\left(\frac{\sqrt{N}\,(D/N - PD)} {\sqrt{PD\,(1-PD)}}\right).
\end{equation}
\item The test for $H_0$ as required by the ECB would typically be performed when
    $D/N > PD$, i.e.\ when there are doubts with regard to the conservatism of
     the PD estimate. Rejection of $H_0$ would then be regarded as `proof' of the estimate
     being aggressive while non-rejection would entail `acquittal' for lack of evidence.
     In case of $1/2 \ge D/N > PD$, it holds that $PD\,(1-PD) < D/N\,(1-D/N)$ such
     that the p-value according to the ECB test is lower than the p-value according to
     \eqref{eq:Xsimple} and \eqref{eq:simple}, i.e.\ the ECB test would reject $H_0$ earlier
     than the simplified version of the test according to Section~\ref{se:TestPD}.
\end{itemize}
	
\section{Numerical examples}
\label{se:example}

The test methods of Section~\ref{se:paired} and the appendices are illustrated
in Section~\ref{se:ExUnitInterval} below with numerical results from
tests on a data set from \citet[][Table~1]{fischer2014statistical}. The test methods
of Section~\ref{se:PD} are illustrated
in Section~\ref{se:ExProb} below with numerical results from tests on 
a data set consisting of simulated data. However, the exposures in the data set
are again from \citet[][Table~1]{fischer2014statistical}. A zip-archive with
the R-scripts and csv-files that were used for computing the results can be downloaded from 
\url{https://www.researchgate.net/profile/Dirk_Tasche}.

\subsection{Example: Tests for variables with values in the unit interval}
\label{se:ExUnitInterval}

\begin{verbatim}
[1] "2020-05-05 20:48:57 CEST"
R Script: PairedDifferences.R
Input data: LGD.csv

Summary of sample distribution:
Sample size: 100 
Sample means:
EqWeighted   Weighted 
   0.02110   -0.09814 
Sample standard deviations:
EqWeighted   Weighted W.adjusted 
    0.3011     0.3186     0.6419 
Three largest weights: 0.07998 0.07994 0.04192 
Sample quantiles:
    10%     25%     50%     75%     90% 
-0.3770 -0.2075  0.0250  0.2700  0.4580 
Weight-adjusted sample quantiles:
       10%        25%        50%        75%        90% 
-0.4474643 -0.0921226  0.0009514  0.0617173  0.3173528 

Random seed: 23 
Bootstrap iterations: 999 


p-values for H0: mean(obs-pred)>=0 vs. H1: mean(obs-pred)<0
                  Eq-weighted Weighted W-adjusted
t-test                 0.7564 0.001403    0.06569
Basic                  0.7240 0.001000    0.07000
Basic normal           0.7583 0.001034    0.06314
Expanded variance      0.6670 0.016000    0.11400
Exp var normal         0.6846 0.015140    0.13474

p-values for H0: mean(obs-pred)<=0 vs. H1: mean(obs-pred)>0
                  Eq-weighted Weighted W-adjusted
t-test                 0.2436   0.9986     0.9343
Basic                  0.2770   1.0000     0.9310
Basic normal           0.2417   0.9990     0.9369
Expanded variance      0.3340   0.9850     0.8870
Exp var normal         0.3154   0.9849     0.8653
\end{verbatim}

\paragraph{Explanations.}\ 
\begin{itemize}
\item Sample means: According to \eqref{eq:Deltaw}. Weights according to \eqref{eq:wLGD} 
with EAD from the column `raw.w' of the data set, and $w_i = 1/100$ in the equally weighted case.
\item Sample standard deviations: First two values according to the square root of the right-hand side of
\eqref{eq:varX}. Third
value also according to \eqref{eq:varX}, but with $\widetilde{\Delta}_i$ from (A3.a) and
equal weights.
\item Weights according to \eqref{eq:wLGD} with EAD from the column `raw.w' of the data set.
\item Sample quantiles: Based on sample $\Delta_1, \ldots, \Delta_{100}$ computed as difference
of columns `obs' and `pred' of the data set.
\item Weight-adjusted sample quantiles: Based on sample $\widetilde{\Delta}_1, \ldots, \widetilde{\Delta}_{100}$
according to (A3.a).
\item t-test results: `Eq-weighted' according to \eqref{eq:pvaluet} and $1-\text{p-value}^\ast$ for the first row
of the t-test results. `Weighted' analogously adapted for the weighted case (but without strong theoretical
foundation). `W-adjusted' like `Eq-weighted' but for the sample 
$\widetilde{\Delta}_1, \ldots, \widetilde{\Delta}_{100}$.
\item `Basic' results: Bootstrapped according to \eqref{eq:bootp} and \eqref{eq:bootpStar}
respectively, with weights and samples like for the t-test rows.
\item `Basic normal' results: Normal approximations according to \eqref{eq:normalp} and
\eqref{eq:normalpStar} respectively, with weights and samples like for the t-test rows.
\item `Expanded variance' results: With weights and samples like for the t-test rows, bootstrapped according to \eqref{eq:bootpLGD} and \eqref{eq:bootpStarLGD} respectively for the first two values, and according to 
(B6.a) and (B6.b) respectively for the third value.
\item `Exp var normal' results: With weights and samples like for the t-test rows, 
normal approximations according to \eqref{eq:normalpLGD} and
\eqref{eq:normalpStarLGD} respectively for the first two values, and according to 
(B7.a) and (B7.b) respectively for the third value.
\end{itemize}
This example demonstrates that
\begin{itemize}
\item test results based on equally weighted means and means with inhomogeneous weights
can lead to contradictory conclusions,
\item variance expansion to capture the individual randomness of single observation-prediction pairs 
can have some impact on the degree of certainty of the test results, by entailing greater p-values, and
\item the two different approaches to account for the weights of the observation-prediction pairs
discussed in this paper can deliver similar but still clearly different results.
\end{itemize}

\subsection{Example: Testing probabilities on inhomogeneous samples}
\label{se:ExProb}

\begin{verbatim}
[1] "2020-05-05 20:50:56 CEST"
R Script: Probabilities.R
Input data: PD.csv

Summary of sample distribution:
Sample size: 100 
Sample means:
EqWeighted   Weighted 
   0.01913    0.06584 
Sample standard deviations:
EqWeighted   Weighted 
    0.3023     0.3367 
Three largest weights: 0.07998 0.07994 0.04192 
Sample quantiles:
       10%        25%        50%        75%        90% 
-0.1803235 -0.0359915 -0.0043570 -0.0005316  0.1322876 

Random seed: 23 
Bootstrap iterations: 999 

p-values for H0: mean(obs-pred)>=0 vs. H1: mean(obs-pred)<0
                  Eq-weighted Weighted
Jeffreys               0.7668       NA
Basic                  0.7390   0.9770
Basic normal           0.7365   0.9747
Expanded variance      0.6525   0.9295
Exp var normal         0.6902   0.9315

p-values for H0: mean(obs-pred)<=0 vs. H1: mean(obs-pred)>0
                  Eq-weighted Weighted
Jeffreys               0.2332       NA
Basic                  0.2620  0.02400
Basic normal           0.2635  0.02526
Expanded variance      0.3475  0.07048
Exp var normal         0.3098  0.06850
\end{verbatim}

\paragraph{Explanations.}\ 
\begin{itemize}
\item See Section~\ref{se:ExUnitInterval} for an explanation of the summary of
the sample distribution.
\item `Jeffreys' results: The `Eq-weighted' value for `H0: mean(obs-pred)$\le$0 vs.~H1: mean(obs-pred)$>$0'
is computed according to \eqref{eq:Jeffreys}. The `Eq-weighted' value for 
`H0: mean(obs-pred)$\ge$0 vs.~H1: mean(obs-pred)$<$0' is $1-\text{p-value}_{\text{Jeffreys}}$. 
No `Weighted' results are computed because there is no obvious `weighted mean'-version of
the binomial Jeffreys test.
\item `Basic' results: Bootstrapped according to \eqref{eq:bootp} and \eqref{eq:bootpStar}
respectively.
\item `Basic normal' results: Normal approximations according to \eqref{eq:normalp} and
\eqref{eq:normalpStar} respectively.
\item `Expanded variance' results: Exact p-values by inverse Fourier transform 
according to \eqref{eq:p} and \eqref{eq:pStar} respectively.
\item `Exp var normal' results: Normal approximations according to \eqref{eq:normalpProb} and
\eqref{eq:normalpStarProb} respectively.
\end{itemize}
This example demonstrates that
\begin{itemize}
\item as mentioned in Section~\ref{se:ECBtest}, the Jeffreys test has a tendency to
earlier reject `H0: mean(obs-pred)$\le$0' than the other tests discussed in Section~\ref{se:PD},
\item test results based on equally weighted means and means with inhomogeneous weights
can lead to different outcomes (no conclusion vs.\ rejection of the null hypothesis), and
\item variance expansion to capture the individual randomness of single observation-prediction pairs 
can have some impact on the degree of certainty of the test results, by entailing greater p-values.
\end{itemize}


\section{Conclusions}
\label{se:concl}

In this paper, we have made suggestions of how to improve on the t-test and the Jeffreys test 
presented in \citet{ECBintructions} for assessing the `preditive ability (or calibration)' 
of credit risk parameters. The improvements refer to 
\begin{itemize}
\item also testing the null hypothesis that the estimated parameter is less than 
or equal to the true parameter in order to be able to `prove' that the estimate is
prudent (or conservative),
\item additionally using exposure- or limit-weighted sample averages in order to 
better inform assessments of estimation (or prediction) prudence, and
\item `variance expansion' in order to account for sample inhomogeneity in terms of
composition (exposures sizes) and riskiness. 
\end{itemize}
The suggested test methods have been illustrated with exemplary test results. R-scripts
with code for the tests are available.

\section*{Acknowledgments}

The author is grateful to two anonymous reviewers whose comments redounded to
significant improvements of the paper.

\section*{Conflicts of interest}

The author declares no conflicts of interest in this paper.


\bibliographystyle{plainnat}
\bibliography{ProvingPredictionPrudence_corr}

\addcontentsline{toc}{section}{References}

\appendix

\section{Appendix: Special cases of the weighted paired difference approach}
\label{se:Cases}

\paragraph{Equal weights in the basic approach.} In this case, the variable of interest is
the ordinary average of the sample $\Delta_1, \ldots, \Delta_n$, as reflected by the fact
that then instead of \eqref{eq:meanX}, it holds that
\begin{equation}
\e[X_\vartheta] \ = \ \frac 1 n \sum_{i=1}^n \Delta_i -\vartheta.
\end{equation}
In the same vein, the algorithms and formulae of Section~\ref{se:basic} can 
be adapted to the equal weights case by replacing all weights $w_i$ and $w_j$ with $1/n$.

\paragraph{Weight-adjusted sample.} In this case, the weights $w_i$ are accounted for by replacing
the sample $\Delta_1, \ldots, \Delta_n$ with the sample $\Delta^\ast_1, \ldots, \Delta^\ast_n$ where 
$\Delta^\ast_i$ is defined by
\begin{equation*}
\Delta^\ast_i \ =\ w_i\,\Delta_i.
\end{equation*}
The adjusted sample $\Delta^\ast_1, \ldots, \Delta^\ast_n$ in turn is treated as in the equal weights case. 
Then, in particular, \eqref{eq:DistrX} for the distribution of $X_\vartheta$ reads
\begin{equation*}
\p[X_\vartheta = \Delta^\ast_i - \vartheta]\ = \ \frac 1 n, \qquad i = 1, \ldots, n.
\end{equation*}
If $\sum_{i=1}^n w_i\,\Delta_i \not= 0$, it follows that
\begin{equation}\label{eq:not}
\e[X_\vartheta] \ = \ \frac 1 n \sum_{i=1}^n \Delta^\ast_i - \vartheta\ \not= \ 
    \sum_{i=1}^n w_i\,\Delta_i - \vartheta.
\end{equation}
As a consequence of \eqref{eq:not}, the adaptation of the algorithms and formulae from
Section~\ref{se:basic} for the weight-adjusted sample case would 
appear somewhat misleading if comparability in magnitude of the values of the
test statistic $\bar{X}_\vartheta$ to its values in the unequal weights case as discussed in
Section~\ref{se:basic} were intended. 

A workaround for this problem is to adjust the sample not only for the weights but also for
the sample size, i.e.\ to define the adjusted sample $\widetilde{\Delta}_1, \ldots, \widetilde{\Delta}_n$ by
\begin{subequations}
\begin{equation}\label{eq:adjCase}
\widetilde{\Delta}_i \ =\ n\,w_i\,\Delta_i.
\end{equation}
Assuming equal weights now means $\p[X_\vartheta = \widetilde{\Delta}_i- \vartheta] = 1/n$ which implies
\begin{align}
\e[X_\vartheta] & \ =\ \frac 1 n \sum_{i=1}^n \widetilde{\Delta}_i - \vartheta\notag\\ 
& \ = \ \sum_{i=1}^n w_i\,\Delta_i - \vartheta, \\
\var[X_\vartheta] & \ =\  \frac 1 n \sum_{i=1}^n \widetilde{\Delta}^2_i - 
    \left(\frac 1 n\sum_{i=1}^n \widetilde{\Delta}\right)^2\notag \\
 & \ = \ n \sum_{i=1}^n w_i^2\,\Delta_i^2 - \left(\sum_{i=1}^n w_i\,\Delta_i\right)^2.   
\end{align}
\end{subequations}
Comparison with \eqref{eq:varX} shows that the variances of $X_\vartheta$ according to the weighting
scheme \eqref{eq:adjCase} and the weighting scheme deployed in Section~\ref{se:basic} differ
by 
\begin{equation*}
\sum_{i=1}^n (n\,w_i-1)\,w_i\,\Delta_i^2,
\end{equation*}
which can be positive or negative.
The algorithms and formulae from Section~\ref{se:basic} can 
be applied to the weight-adjusted sample case as specified by \eqref{eq:adjCase} 
and $\p[X_\vartheta = \widetilde{\Delta}_i - \vartheta] = 1/n$
if the following two modifications are taken into account in the given order:
\begin{itemize}
\item Replace the value of $\Delta_i$ by the value of 
    $\widetilde{\Delta}_i \ =\ n\,w_i\,\Delta_i$ for $i = 1, \ldots, n$.
\item Replace all remaining appearances of the weights $w_i$ by $1/n$.
\end{itemize}

Note that the weight-adjustment \eqref{eq:adjCase} can also be deployed for samples
with more special structure like the ones considered in Section~\ref{se:LGD.CCF}
and Appendix~\ref{se:EAD} below. There is no guarantee, however, that
adjustment \eqref{eq:adjCase} would preserve the `values in the unit interval'
constraint of Section~\ref{se:LGD.CCF}. There is no such preservation issue 
with regard to Appendix~\ref{se:EAD}.

\section{Appendix: Tests for non-negative variables}
\label{se:EAD}

In contrast to LGD and CCF which by definition are variables with values in the unit interval, EAD 
in principle may take any non-negative value. This requires some modifications
in order to adapt the approach from Section~\ref{se:LGD.CCF} to the assessment of
EAD estimates.

\paragraph{Starting point.}
\begin{itemize}
\item A sample of paired observations $(h_1, \eta_1), \ldots, (h_n, \eta_n)$, with 
predicted EADs $0 < \eta_i < \infty$ and 
realised exposures $0 \le h_i < \infty$.
\item Weights $0 < w_i < 1$, $i = 1, \ldots, n$, with $\sum_{i=1}^n w_i = 1$,
\item Weighted average observed EAD $h_w = \sum_{i=1}^n w_i\,h_i$ and
weighted average EAD prediction $\eta_w = \sum_{i=1}^n w_i\,\eta_i$.
\end{itemize}

\paragraph{Interpretation in the context of EAD back-testing.} 
\begin{itemize}
\item A sample of $n$ defaulted credit facilities / loans is analysed.
\item The EAD $\eta_i$ is an estimate of loan $i$'s exposure at the moment of the default, measured
in currency units.
\item The realized exposure $h_i$ shows the loan $i$'s exposure at the time of default. 
\item The weight $w_i$ reflects the relative importance of observation $i$. In the case of direct 
EAD predictions, one might choose $w_i$ according to \eqref{eq:wCCF}.
\item Define $\Delta_i = h_i - \eta_i$, $i = 1, \ldots, n$. If $|\Delta_i| \approx 0$ then $\eta_i$ 
is a good EAD prediction.
If $|\Delta_i|$ is large then $\eta_i$ is a poor EAD prediction.
\end{itemize}

\paragraph{Goal.} We want to use the observed weighted average difference / residual 
$\Delta_w = \sum_{i=1}^n w_i\,\Delta_i = h_w - \eta_w$ to 
assess the quality of the calibration of the model / approach for the $\eta_i$ to
predict the realised exposures $h_i$. Again we want to answer the following two questions:
\begin{itemize}
\item If $\Delta_w <0$, how safe is the conclusion that the observed (realised) values are
on weighted average less than the predictions, i.e.\ the predictions are prudent / conservative?
\item If $\Delta_w > 0$, how safe is the conclusion that the observed (realised) values are
on weighted average greater than the predictions, i.e.\ the predictions are aggressive?
\end{itemize}
The safety of such conclusions is measured by p-values which provide error probabilities for the
conclusions to be wrong. The lower the p-value, the more likely the conclusion is right.

In order to be able to examine the specific properties of 
the sample and $\Delta_w$ with statistical methods, 
we have to make the assumption that the sample was generated with
some random mechanism. This mechanism is described in the following modification
of Assumption~\ref{as:LGD}.

\begin{subequations}
\begin{assumption}\label{as:EAD}
The sample $\Delta_1, \ldots, \Delta_n$ consists of independent realisations of a random variable
$X_\vartheta$ with distribution given by
\begin{equation}
X_\vartheta \ = \ h_I - Y_\vartheta,
\end{equation}
where $I$ is a random variable with values in $\{1, \ldots, n\}$ and $\p[I=i] = w_i$, $i=1, \ldots, n$.
$Y_\vartheta$ is a gamma$(\alpha_i,\beta_i)$-distributed random variable\footnote{%
See \citet[][Section 3.3]{Casella&Berger} for a definition of the gamma-distribution.} conditional
on $I=i$ for $i=1, \ldots, n$. 
The parameters $\alpha_i$ and $\beta_i$ of the gamma-distribution depend on 
the unknown parameter $0 < \vartheta < \infty$ by
\begin{equation}\label{eq:gammapars}
\begin{split}
\alpha_i & \ = \ \frac{\vartheta_i}{v}, \qquad\text{and}\\
\beta_i & \ = \ v.
\end{split}
\end{equation}
In \eqref{eq:gammapars}, the constant $0 < v < \infty$ is the same for all $i$. The $\vartheta_i$ are
determined by
\begin{equation}\label{eq:theta.gamma}
\vartheta_i \ = \ \eta_i\,\frac{\vartheta}{\eta_w}.
\end{equation}
\end{assumption}
\end{subequations}

Note that Assumption~\ref{as:EAD} describes a method for
recalibration of the EAD estimates $\eta_1, \ldots, \eta_n$
to match targets $\vartheta$ with the weighted average of the $\vartheta_i$.
By definition of $Y_\vartheta$, it holds that $\e[Y_\vartheta\,|\,I=i] = \vartheta_i$. 

The constant $v$ specifies the variance of $Y_\vartheta$ conditional on $I=i$ as multiple 
of its expected value $\vartheta_i$, i.e.\ it holds that
\begin{equation}
\var[Y_\vartheta\,|\,I=i] \ = \ v\,\vartheta_i, \qquad i = 1, \ldots, n.
\end{equation}
The constant $v$ must be pre-defined or separately estimated. We suggest estimating it
from the sample $h_1, \ldots, h_n$ as 
\begin{equation}
\hat{v}\ =\ \frac{\sum_{i=1}^n w_i\,h_i^2 - h_w^2}{h_w}.
\end{equation}

\begin{proposition}\label{pr:momentsEAD}
For $X_\vartheta$ as described in Assumption~\ref{as:EAD}, the expected value and 
the variance are given by
\begin{subequations}
\begin{align}
\e[X_\vartheta] & = h_w - \vartheta, \text{\ and}\label{eq:meanEAD}\\
\var[X_\vartheta]  & = \sum_{i=1}^n w_i\,(h_i-\vartheta_i)^2 - (h_w - \vartheta)^2 + 
    v\,\vartheta.\label{eq:varEAD}
\end{align}
\end{subequations}
\end{proposition}

\textbf{Proof.} For deriving the formula for $\var[X_\vartheta]$, make use of
the well-known variance decomposition \\[1ex]
\hspace*{3cm}$\displaystyle\var[X_\vartheta] = \e\bigl[\var[X_\vartheta\,|\,I]\bigr] +
\var\bigl[\e[X_\vartheta\,|\,I]\bigr]$.\hfill{\qed}

Like in \eqref{eq:varLGD}, the variance of $X_\vartheta$ as shown in \eqref{eq:varEAD}
depends on the parameter $\vartheta$ and has an additional component $v \,\vartheta$
which reflects the potentially different variances of the exposures at default 
in an inhomogeneous portfolio.

By Assumption~\ref{as:EAD} and Proposition~\ref{pr:momentsEAD}, the questions on the
safety of conclusions from the sign of $\Delta_w$ again can be translated into 
hypotheses on the value of the parameter $\vartheta$:
\begin{itemize}
\item If $\Delta_w < 0$, can we conclude that $H_0: \vartheta \le h_w$ is false and
$H_1: \vartheta > h_w \Leftrightarrow \e[X_\vartheta] < 0$ is true?
\item If $\Delta_w > 0$, can we conclude that $H^\ast_0: \vartheta \ge h_w$ is false and
$H^\ast_1: \vartheta < h_w \Leftrightarrow \e[X_\vartheta]> 0$ is true?
\end{itemize}

If we assume that the sample $\Delta_1, \ldots, \Delta_n$ was generated by independent
realisations of $X_\vartheta$ then the distribution of the sample mean is different from
the distribution of $X_\vartheta$, as shown in the following corollary to Proposition~\ref{pr:momentsEAD}.

\begin{corollary}\label{co:meansEAD}
Let $X_{1, \vartheta}, \ldots, X_{n, \vartheta}$ be independent 
and identically distributed copies of $X_\vartheta$ 
as in Assumption~\ref{as:EAD} and 
define $\bar{X}_\vartheta = \frac{1}{n} \sum_{i=1}^n X_{i, \vartheta}$. Then for 
the mean and variance of $\bar{X}_\vartheta$, it holds that
\begin{subequations}
\begin{align}
\e[\bar{X}_\vartheta] & = h_w - \vartheta.\label{eq:meanEADbar}\\
\var[\bar{X}_\vartheta]  & = \frac{1}{n} 
    \left(\sum_{i=1}^n w_i\,(h_i-\vartheta_i)^2 - (h_w - \vartheta)^2 + 
    v \,\vartheta\right). \label{eq:varEADbar}
\end{align}
\end{subequations}
\end{corollary}
In the following, we use $\bar{X}_\vartheta$ as the test statistic and interpret 
$\Delta_w = h_w - \eta_w$ as its observed value.

\begin{proposition}\label{pr:EAD}
In the setting of Assumption~\ref{as:EAD} and Corollary~\ref{co:meansEAD}, 
$\vartheta \le \widehat{\vartheta}$ implies that
\begin{equation*}
\p[\bar{X}_\vartheta \le x]\ \le \ \p[\bar{X}_{\widehat{\vartheta}} \le x],\qquad \text{for all }x\in\mathbb{R}.
\end{equation*}
\end{proposition}
\textbf{Proof.} Same as the proof of Proposition~\ref{pr:LGD}.

\begin{subequations}
\paragraph{Bootstrap test.} Generate a Monte Carlo sample
$\bar{x}_1, \ldots, \bar{x}_R$ from $X_\vartheta$ with $\vartheta = h_w$ as follows:
\begin{itemize}
\item For $j=1, \ldots, R$: $\bar{x}_j$ is the equally weighted mean of $n$ 
independent draws from the distribution of $X_{\vartheta}$ as given by Assumption~\ref{as:EAD},
with $\vartheta = h_w$. 
\item $\bar{x}_1, \ldots, \bar{x}_R$ are realisations of independent, 
identically distributed random variables,
\end{itemize}
Then a bootstrap p-value for the test of $H_0: \vartheta \le h_w$ 
against $H_1: \vartheta > h_w$ can be calculated as
\begin{equation}\label{eq:bootpEAD}
\text{p-value} \ = \ \frac{1 + \#\bigl\{i: i \in\{1, \ldots, n\}, 
    \bar{x}_i \le h_w- \eta_w\bigr\}}{R+1}.
\end{equation}
A bootstrap p-value for the test of $H^\ast_0: \vartheta \ge h_w$ 
against $H^\ast_1: \vartheta < h_w$ is given by
\begin{equation}\label{eq:bootpStarEAD}
\text{p-value}^\ast \ = \ \frac{1 + \#\bigl\{i: i \in\{1, \ldots, n\}, 
    \bar{x}_i \ge h_w- \eta_w\bigr\}}{R+1}.
\end{equation}
\end{subequations}

\paragraph{Rationale.} Same as the rationale for \eqref{eq:bootpLGD} and \eqref{eq:bootpStarLGD}.

\begin{subequations}
\paragraph{Normal approximate test.} By Corollary~\ref{co:meansEAD}, 
we find that the distribution of $\bar{X}_{h_w}$ can be approximated by a normal distribution with 
mean 0 and variance as shown  on the right-hand side of \eqref{eq:varEADbar} 
with $\vartheta=h_w$. With  $x=h_w-\eta_w$,
one obtains for the approximate p-value of $H_0: \vartheta \le h_w$ against $H_1: \vartheta > h_w$:
\begin{align}
\text{p-value} &\ = \ \p[\bar{X}_{h_w} \le x] \notag\\
    & \ \approx\ \Phi\left(\frac{\sqrt{n}\,(h_w - \eta_w)}
    {\sqrt{\sum_{i=1}^n w_i\,(h_i-\widehat{\vartheta}_i)^2  + v\,h_w}}\right),\label{eq:normalpEAD}
\end{align}
with $\widehat{\vartheta}_i = \eta_i\,\frac{h_w}{\eta_w}$ as in
Assumption~\ref{as:EAD}.
The same reasoning gives for the normal approximate p-value of 
$H^\ast_0: \vartheta \ge h_w$ against $H^\ast_1: \vartheta < h_w$:
\begin{equation}\label{eq:normalpStarEAD}
\text{p-value}^\ast  \ \approx\ 1 - \Phi\left(\frac{\sqrt{n}\,(h_w - \eta_w)}
    {\sqrt{\sum_{i=1}^n w_i\,(h_i-\widehat{\vartheta}_i)^2  + v\,h_w}}\right).
\end{equation}
\end{subequations}

\end{document}